# A Proposed Hybrid Recommender System for Tourism Industry in Iraq Using Evolutionary Apriori and K-means Algorithms

Bryar A. Hassan*[1,2], Alla A. Hassan[3], Joan Lu[4], Aram M. Ahmed[1], Tarik A. Rashid[1]

[1]Computer Science and Engineering Department, School of Science and Engineering, University of Kurdistan Hewler, Erbil, Iraq.

[2]Department of Computer Science, College of Science, Charmo University, 46023 Chamchamal/Sulaimani, KRI, Iraq.

[3]Department of Database Technology, Computer Science Institute, Sulaimani Polytechnic University, Sulaimani, Iraq.

[4]Department of Computer Science, School of Computing and Engineering, University of Huddersfield, Queensgate, Huddersfield HD1 3DH, UK

Email (Corresponding): bryar.ahmad@ukh.edu.krd

**Abstract**

The rapid proliferation of tourism data across sectors, including accommodations, cultural sites, and events, has made it increasingly challenging for travelers to identify relevant and personalized recommendations. While traditional recommender systems such as collaborative, content-based, and context-aware systems offer partial solutions, they often struggle with issues like data sparsity and overspecialization. This study proposes a novel hybrid recommender system that combines evolutionary Apriori and K-means clustering algorithms to improve recommendation accuracy and efficiency in the tourism domain. Designed specifically to address the diverse and dynamic tourism landscape in Iraq, the system provides personalized recommendations and clusters of tourist destinations tailored to user preferences and contextual information. To evaluate the system's performance, experiments were conducted on an augmented dataset representative of Iraq's tourism activity, comparing the proposed system with existing methods. Results indicate that the proposed hybrid system significantly reduces execution time by 27-56% and space consumption by 24-31%, while achieving consistently lower Root Mean Square Error (RMSE) and Mean Absolute Error (MAE) values, thereby enhancing prediction accuracy. This approach offers a scalable, context-aware framework that is well-suited for application in regions where tourism data is limited, such as Iraq, ultimately advancing tourism recommender systems by addressing their limitations in complex and data-scarce environments.

## 1. Introduction

The rapid growth of tourism and the widespread availability of tourism-related data across sectors, such as hotels, heritage sites, transportation, and tourist events have intensified the need for efficient recommendation systems. With the advent of digital platforms and travel applications, travelers face an overwhelming array of options, often finding it challenging to locate relevant, personalized recommendations amidst extensive information. Recommender systems, which filter and rank data to provide the most relevant suggestions, have emerged as powerful tools to enhance trip planning and streamline user choices in the tourism sector.

In recent years, various types of recommender systems have been developed, including collaborative, content-based, and context-aware systems. Each approach has demonstrated specific strengths, such as collaborative filtering's effectiveness in capturing similar user preferences and content-based filtering's ability to recommend items based on users' historical interactions. However, these methods also face limitations. Collaborative filtering can suffer from sparsity issues when users or items have minimal interactions, while content-based approaches may lead to overspecialization by repeatedly recommending similar items. In response to these challenges, hybrid recommender systems that integrate multiple approaches have gained traction, aiming to leverage the strengths of different methodologies while mitigating their weaknesses.

The underlying premise involves utilizing the user's interests, which are gathered throughout their browsing activities, as inputs to forecast the level of interest that the user may have towards a specific item. There exist a multitude of methodologies employed to ascertain various levels of valuation. The categorization of these entities is conventionally determined by scholars in the field of literature, who classify them into several groups based on the specific sources of information utilized [1][2]. One of the methodologies employed in this study is the utilization of ratings provided by a group of individuals for a certain collection of products. The approach involves suggesting goods to a specific user based on the high ratings given by other users with similar tastes. This technique is commonly referred to as collaborative filtering [3]. Social recommender systems have been developed by incorporating social information alongside the rating matrix, leveraging the emergence of social networks [4]. These systems involve the computation of similarities between the target user and their social environment, based on a certain measure. In recent years, there has been a growing trend in research to incorporate contextual information, such as location and weather conditions, into recommender systems. This is primarily driven by the recognition of the significant role that contextual factors play in the effectiveness of these systems [5][6]. As stated in reference [7], the term "context" is defined as "any information that can be utilized to describe the circumstances surrounding an entity." An entity might encompass individuals, locations, or items that are deemed significant within a certain context. Within the realm of tourism, objects can encompass a variety of categories, including but not limited to monuments, parks, museums, and other such entities.

This study proposes a hybrid recommender system that combines evolutionary Apriori association rule mining with K-means clustering to enhance recommendation accuracy and efficiency in the tourism sector. This approach was chosen to address the unique demands of tourism in Iraq, where diverse tourist attractions and fluctuating user preferences require a flexible, context-aware recommendation framework. The proposed system not only aims to provide personalized recommendations but also organizes tourism options into meaningful clusters, thus enhancing both the relevance and computational efficiency of the recommendations.

**1.1. Research questions and objectives**

The key research questions guiding this study are as follows:

- How can a hybrid recommender system improve the accuracy and efficiency of tourism recommendations, particularly in the context of Iraq?
- What are the benefits of integrating evolutionary Apriori and K-means algorithms in a tourism recommender system?

To answer these questions, the study sets out with the following objectives:

- To develop a hybrid recommender system that effectively integrates association rule mining and clustering, providing more personalized and context-aware recommendations.
- To evaluate the system's performance in terms of accuracy, computational efficiency, and resource consumption, using a dataset that simulates the patterns of tourist activity in Iraq.
- To contribute to the advancement of tourism recommender systems by offering a framework that can adapt to diverse user preferences and a wide range of attractions.

**1.2. Contributions and novelty**

The primary contributions of this study are as follows:

- Development of a Hybrid System: The proposed system combines evolutionary Apriori for association rule mining and K-means clustering to group tourist locations based on behavioral patterns. This integration enhances recommendation accuracy by aligning user preferences with grouped attractions, ensuring relevance in diverse and dynamic contexts.
- Efficiency in Recommendation Processing: By employing a hybrid approach, the system reduces both execution time and space consumption, making it scalable for large datasets. This computational efficiency is particularly valuable in real-time applications where response speed is critical.
- Application to Iraqi Tourism: The system is tailored to the unique context of Iraq, where a lack of real-time digital tourism data presents challenges for traditional recommendation approaches. This study provides a model that can adapt to data constraints and meet the specific needs of the Iraqi tourism sector.

The remainder of this paper is structured as follows: Section 2 introduces the algorithms related to research background of this study. Section 3 provides an overview of related work, Section 4 describes the proposed system architecture, Section 5 outlines the experimental setup, and Section 6 presents the results and evaluation of the system. The paper concludes with insights on the system's implications and future research directions.

## 2. Research background

This section introduces the algorithms and research backgrounds needed for this study.

### 2.1. Apriori association rule mining

An association rule is a model that outlines a procedure for analyzing the co-occurrence of events and specific kinds of data linkages [8]. Association rules identify correlations between occurrences and show how likely it is for them to occur. This model's primary goal is to highlight intriguing data correlations. These days, this technique is applied in numerous disciplines, including bioinformatics, continuous manufacturing, intrusion detection, and Web usage mining. The algorithms for generating rules are Apriori [9], Eclat and FP-growth, SETM, Partition, RARM - Rapid Association Rule Mining, and CHARM. In this paper, we utilize the Apriori algorithm.

The Apriori algorithm operates as follows:

1. Association rule mining is well-defined as: $I = \{i_1, i_2,...,i_n\}$ and $D = \{t_1, t_2,..., t_m\}$, $I$ where:
- $I$ is a set of two attributes named items,
- $D$ is a set of transactions named database,
- Every transaction in $D$ has a transaction identity and includes a subset of the item in $I$.

2. A rule can be demonstrated as the following form: $X \Rightarrow Y$ $X, Y \subseteq I$. A rule is presented only between a set and a single item, $X \Rightarrow i_j$ for $i_j \in I$. Every rule is composed by two different sets of items, also called itemsets, X and Y, where X is called antecedent and Y consequent.

3. Constraints on many metrics of relevance and interest are employed to effectively choose captivating rules from the comprehensive collection of all potential rules. The most widely recognized limitations are the minimal criteria imposed on support and confidence.

4. X, Y be itemsets, $X \Rightarrow Y$ an association rule and T a set of transactions of a given database. Support, confidence and lift are three measures of the relationship or rule [10]. From these, three commonly used measures of association can be estimated.

5. The Equation (1) provides support, which serves as a measure of the frequency with which the itemset occurs within the given data set. The concept of support, denoted as supp (X, T), refers to the relative frequency of transactions in a given dataset, T, that contain a specific itemset, X.

$$Supp(X) = \frac{|\{t \in T; X \sqsubseteq t\}|}{|T|} \qquad (1)$$

6. Confidence as denoted in Equation (2) serves as a metric to quantify the frequency with which the rule has been seen to hold true. The confidence measure of a rule, denoted as $X \Rightarrow Y$, in relation to a given collection of transactions, T, is determined by calculating the proportion of transactions that contain both X and Y. Confidence is commonly understood as a psychological state characterized by a belief in one's abilities, skills, and judgments, which in turn influences one's behavior and decision-making processes.

$$Conf\ (X \Rightarrow Y) = \frac{Supp\ (X \cup Y)}{Supp\ (X)} \qquad (2)$$

7. The lift rule is defined as presented in Equation (3).

$$Lift\ (X \Rightarrow Y) = \frac{Supp\ (X \cap Y)}{Supp\ (X) * Supp\ (Y)} \qquad (3)$$

Alternatively, the ratio can be calculated by dividing the observed support by the expected support under the assumption of independence between variables X and Y.

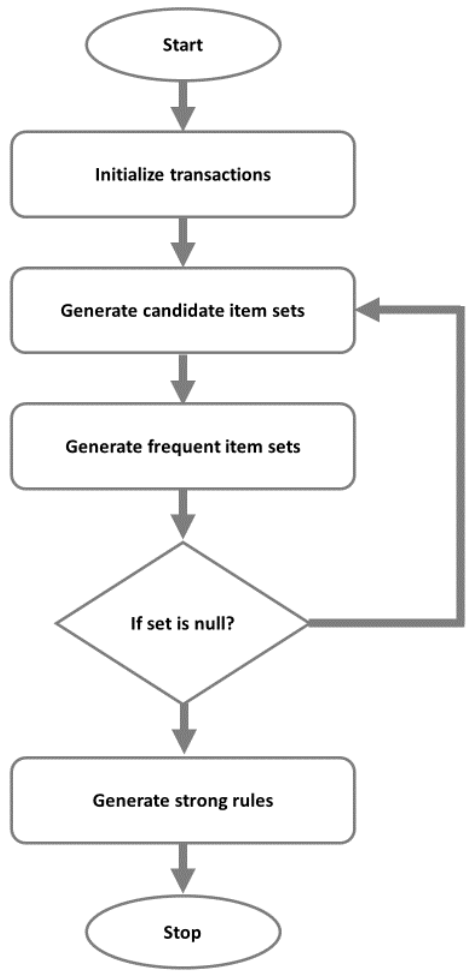

Figure (1): The flowchart of Apriori algorithm

## 2.2. K-Means clustering algorithm

Cluster analysis is a methodological approach that involves the partitioning of data into distinct groups. In contrast to the predetermined categorization approach, the classifications are not pre-established. The allocation of data into groups and clusters, as well as the determination of the number of distinct groups, is based on the similarity of the available data. Cluster analysis is a widely employed technique utilized in various disciplines, including biology, medicine, anthropology, and marketing. The steps of K-means are as the following:

1. Select a value for the total number of clusters, denoted as K.
2. The initial cluster centers are determined by randomly selecting K datapoints from the dataset.
3. In order to allocate the remaining data points to the closest cluster center, the Euclidean distance metric is employed.
4. Each cluster comprises instances that are used to determine the new mean for each cluster.
5. Based on the preceding iteration, if the newly calculated mean values closely resemble the existing mean values, the process is concluded. Furthermore, it is necessary to iterate steps 3-5 for each new set of mean values.

K-means is an unsupervised learning algorithm, meaning it does not require labeled data. It is widely used in various applications such as image segmentation, customer segmentation, anomaly detection, and more. However, it has some limitations, including sensitivity to the initial placement of centroids and the assumption that clusters are spherical and equally sized, which may not always hold true for real-world data. Researchers have developed variations and improvements of K-means to address some of these limitations.

## 3. Related work

The existing methods for recommending tourism options possess several novel elements and can be categorized in diverse manners, contingent upon their approach to analyzing user data and refining the list of available choices [11].

### 3.1. Collaborative recommender system

The objective of this technique is to provide users with recommendations for destinations they have not yet visited but may find appealing. These recommendations are generated by analyzing the habits and preferences of other users with similar profiles [12]. The similarity in taste between two users is determined by comparing their respective rating histories. The VISIT system uses sentiment analysis techniques, namely employing the Alchemy Application Programming Interface (API) [13], to conduct an analysis of news pertaining to a specific attraction on social media platforms such as Twitter and Facebook. This analysis aims to ascertain if users express favorable or negative sentiments towards the attraction. The system's interface utilizes the colors green and red to visually represent information, enabling users to readily discern the current popularity of certain locations among visitors. Nonetheless, this methodology presents challenges in catering to the preferences of visitors. It becomes exceedingly arduous to align users' journey histories, ascertaining ratings, and identifying two individuals who have embarked on identical trips, encompassing comparable durations, destinations of interest, and experiences.

**3.2. Content-based recommender system**

Content-based systems rely on analyzing the similarities in content between objects that users have previously accessed or are currently examining, to provide recommendations to potential visitors [14][15]. The concept of content-based filtering is widely recognized and extensively employed in tourism recommendation systems [16][17]. The content-based recommendation technique for cultural heritage (both tangible and intangible) is outlined in the system described [6]. This approach involves the selection of resources by considering user preferences and item information, the ordering of items through the utilization of multi-criteria user feedback, and the enhancement of the suggested collection of items by incorporating semantic linkages between them. One inherent constraint of content-based filtering pertains to the requirement of possessing a comprehensive and diverse depiction of the items' content. However, this poses a challenge when applied to tourist objects because of their vast scope and heterogeneous nature. In addition, it is worth noting that this particular system often encounters the issue of overspecialization. For instance, the fact that a tourist derives pleasure from attending an event or concert during a journey does not necessarily imply a desire to experience it repeatedly. Nevertheless, employing a content-based methodology, the system will recommend that the individual revisit the same location for a second time, engaging in a similar event, regardless of whether it is organized or not. However, it is possible that the individual may have a greater inclination towards exploring activities that were not previously encountered during their previous visit.

**3.3. Context-aware recommender system**

Recommender systems are said to as context-sensitive when they incorporate contextual information in their computations to anticipate human preferences [18]. The predominant contextual factors employed in tourism recommendation systems encompass geolocation, meteorological conditions, visitation history, and weather forecasts. In contemporary times, there is a prevalent presence and use of mobile devices that are interconnected with the Internet. These devices, commonly referred to as "connected objects," have become extensively accessible and serve the purpose of capturing and delivering a substantial amount of data. This data has the potential to enhance the existing context and its diverse manifestations. The authors of the study [19] introduced a context-sensitive recommender system and provided a comprehensive explanation of the notion of context using a meta-model. A case study with practical applications was undertaken in the urban setting of Tangier. The system that has been developed consists of three primary modules: Context, which encompasses the user's profile, spatio-temporal and environmental information (such as location, device characteristics, and external physical environment like weather conditions), as well as information gathered about the community from social network applications like Facebook and Twitter. The second module is the tourism content repository, which stores data related to tourism services. Lastly, there is the recommender system module.

### 3.4. Hybrid recommender system

Numerous research projects have focused on the integration of various approaches to address the limitations of individual techniques and use their respective advantages [20]. In a previous study, the authors of this research [21] introduced a novel framework for a recommender system designed to cater to both individual users and groups. This recommender system considers several factors such as the attributes of artworks, the contextual information of all users, and the social affinity that exists among these users. The system is an amalgamation of three distinct methodologies: the content-based approach, the social approach, and the context-based approach.

It is important to acknowledge that a significant majority, specifically 90%, of the existing solutions primarily focus on a singular category of items, such as hotels, museums, and tourist sites. These solutions primarily offer information pertaining to tourist services, which is typically input into the system either by the administrator or by experts. The objective of these solutions is to enhance the overall travel experience. Furthermore, it is worth noting that most of these solutions employ a singular approach, with content-based approaches being the most prevalent. There is a necessity for a hybrid recommender system that serves the purpose of not only collecting recommendation techniques but also presenting various tourism materials inside a unified architecture.

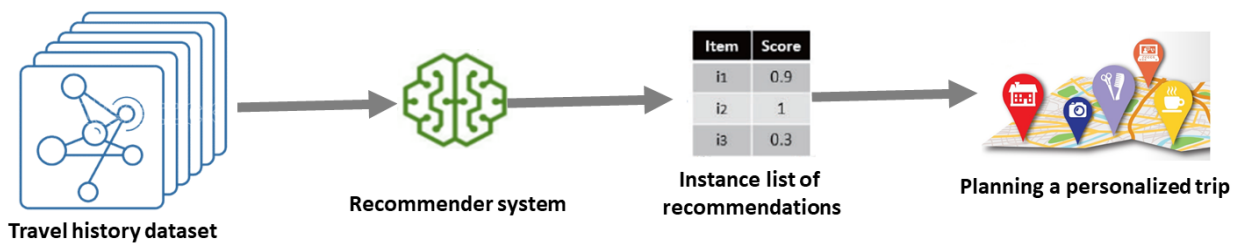

**Figure (2): The current recommender system for tourism purposes**

Table (1) summarizes the related work of the existing methods for recommending tourism options.

Table (1): Summary of the related work of the existing methods for recommending tourism options.

| Reference, year | Type of Recommender System | Objective | Methodology | Limitations | Contribution to Literature |
|---|---|---|---|---|---|
| [6], 2013 | Context-aware | Develop context-sensitive recommendations for tourism | Utilizes user profiles combined with contextual and environmental data | Limited personalization for group recommendations | Introduced real-time context adaptation for tourism recommendations |
| [13], 2013 | Collaborative | Analysing social media sentiment on attractions, showing popularity with colour codes but struggles to personalize recommendations. | Conducting an analysis of news pertaining to a specific attraction on social media platforms. | this methodology presents challenges in catering to the preferences of visitors. | Context-aware intelligent recommendation system for tourism |
| [11], 2014 | Content-based | Improve recommendation for cultural heritage sites | Semantic content filtering designed specifically for cultural recommendations | Overspecialization in repeated activities | Added semantic linkages for enhanced recommendations in cultural heritage |
| [12], 2016 | Collaborative | Provide travel recommendations for groups | Collaborative filtering enhanced with rating predictions | Difficulty in aligning group preferences | Improved collaborative filtering techniques for group-based travel planning |
| [21], 2017 | Social-based | Enhance cultural heritage recommendations using social data | Social recommendation using social networks for cultural data integration | Limited to cultural recommendations, lacks broader tourism application | Incorporated social networks into cultural recommendations |
| [19], 2018 | Context-aware | Introduce smart tourism recommendations applied to Tangier | Employs a meta-model for context-aware recommendations in tourism | Tested only in a specific city (Tangier) context | Developed a model integrating real-time context for enhanced user experience |
| [17], 2020 | Comparative hybrid | Compare hybrid approaches for improved recommendation accuracy | Combines content and collaborative filtering approaches in a comparative study | Focuses on methodological comparison, less on practical application | Showed a comparative analysis of hybrid recommendation methods for tourism |
| [22], 2023 | Group-based | Explore group recommendations based on personality traits and travel motivations | Personality trait analysis for group recommendations | Limited scalability in large, diverse groups | Provides insight into group-based recommendation based on personality in tourism |
| [23], 2022 | Survey | Overview of current methodologies and future directions in tourism recommendations | Survey of current recommender methodologies and emerging trends | General overview; lacks in-depth application analysis | Offers a comprehensive survey and future research suggestions in tourism recommender systems |
| [24], 2023 | Point-of-interest | Enhance accuracy of POI recommendations | Utilizes heterogeneous data sources to increase | Challenges in integrating and processing diverse data sources | Advances POI recommendation accuracy through |

|  |  | by using heterogeneous data | recommendation relevance |  | diverse data integration |

## 4. The proposed recommender system

The present study entails the proposition of a novel architectural method for tourism recommendation systems. The architecture presented in this study is founded upon a hybrid recommendation strategy, with the primary objective of enhancing user accessibility to tourism resources inside information retrieval systems. An additional element of novelty in this architectural design is its capacity to beyond a mere compilation of suggested tourist destinations, instead functioning as a comprehensive planner that endeavors to construct an intricate and comprehensive itinerary for a multi-day excursion. The client will be provided with a comprehensive range of tourism materials that precisely align with their own requirements and preferences.

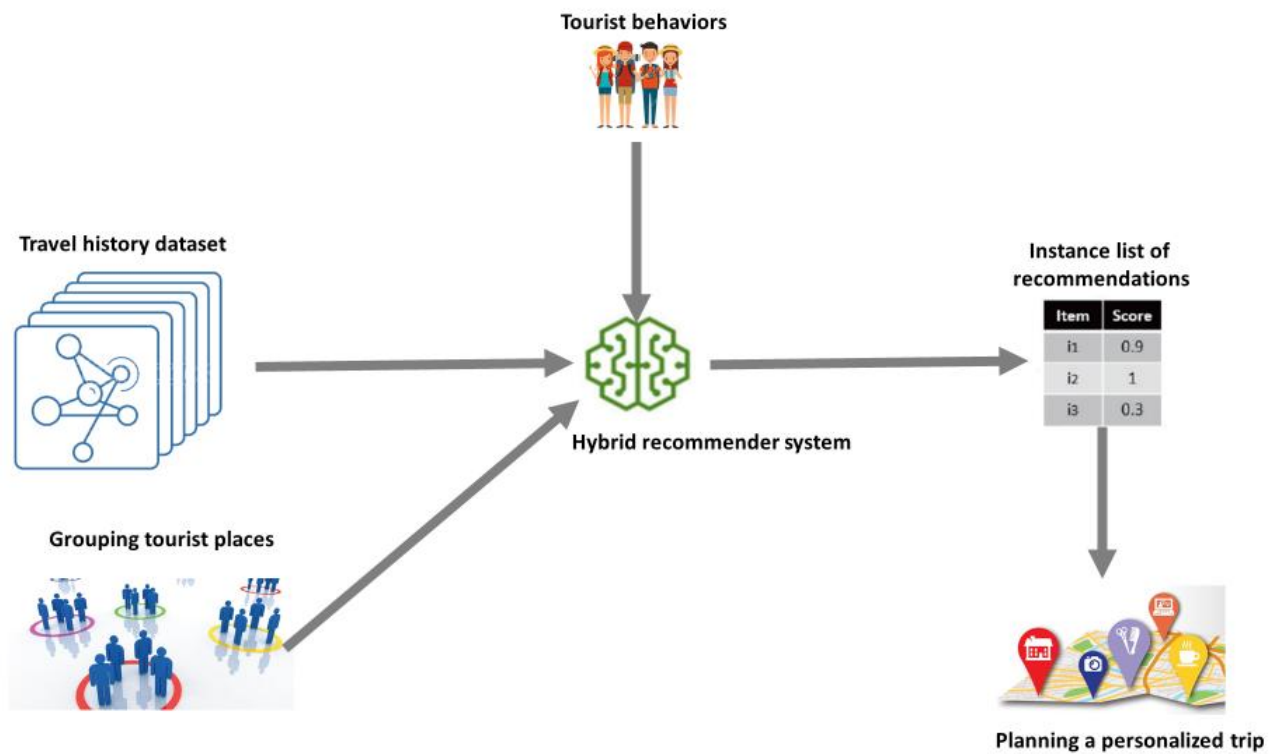

Figure (3): The proposed hybrid recommender system for tourism purposes

In this study, the proposed system architecture is divided into five primary modules, as illustrated in Figure (3).

**1. Data record of visitors:** A set of data is used to train the proposed hybrid recommender system to be able to make more accurate recommendations accordingly.

**2. Tourist behavior:** It includes specific data that can be utilized to ascertain user preferences on the items, such as ratings, age, social information and current geographical location of the visitors.

**3. Grouping tourist places:** A contextual meta-model incorporates the utilization of the k-means algorithm to make the tourist places into several groups. There are several variables that play a role in manipulating context, including temporal elements, spatial factors, distance between two points, travel routes, and the trip history of tourists. These factors are taken into consideration in order to provide a tailored group of recommendations.

**4. Hybrid recommender system:** The output of the module is a collection of items, each accompanied by a corresponding degree of appreciation that represents the extent to which the target user values each item.

**5. The trip planner:** This module employs a selection process to identify objects that are deemed pertinent to the user, and subsequently applies operational research methodologies to establish correlations among these selections, resulting in the creation of a travel itinerary.

The proposed hybrid recommender system offers several advantages over traditional recommender approaches in tourism. By integrating the evolutionary Apriori algorithm and K-means clustering, the system enhances recommendation accuracy through improved association rule mining and clustering, which effectively groups tourist sites based on user preferences and behavioral patterns. This dual approach allows the system to reduce processing time and space consumption, making it computationally efficient. Additionally, the hybrid structure combines the strengths of collaborative and content-based filtering, enabling more relevant, context-aware recommendations tailored to individual tourists. These features make the system particularly suitable for application in tourism domains with diverse datasets and dynamic user preferences, such as those found in Iraq's tourism industry.

## 5. Experiment

To test the suggested recommender systems against the current one, both methods are evaluated on a dataset based on their time execution and space consumption across different minimal support values. The proposed model is analyzed and validated with the existing method using the C++ program. Furthermore, the datasets are initially divided into training and testing, where 70% of datasets are utilized for training, and the remaining 30% are being used for testing. Therefore, this section integrated with the presentation of dataset and the initial parameter tuning for the proposed method.

### 5.1. Dataset overview

The data utilized in this research originates from Kaggle's data repository and comprises the top-rated tourist destinations in Iraq's eighteen provinces [25] . The key components of the dataset consist of ID, the city's name, the name of the place, as well as their respective latitude and longitude coordinates. Additionally, each tourism spot is categorized with a corresponding symbol denoting its type. Table (2) presents the types of Iraqi tourism destinations.

Table (2): The types of Iraqi tourism destinations.

| # | Type | Symbol | Meaning values |
|---|------|--------|----------------|
| 1 | Adventure | A | |
| 2 | Culture | C | |
| 3 | Environmental | E | |
| 4 | Health | H | |
| 5 | Nature | N | 1: Active; Blank: Inactive |
| 6 | Religious | R | |
| 7 | Sport | SP | |
| 8 | Shopping | SH | |
| 9 | Business | B | |
| 10 | Leisure | L | |

**5.2. Data augmentation process**

To provide a robust dataset for training (70%) and testing (20%) the recommendation system, we augmented the initial dataset to create a total of 10,000 records. This augmentation involved synthesizing data for 5,000 individual visitors, each with varying patterns of engagement with 232 distinct tourist attractions. Visitor attributes, including age, gender, preferences, and current location, were generated using a combination of historical trends and probabilistic sampling to emulate realistic tourist behavior. The dataset was augmented with Python, and the generated data was validated by comparing distribution trends against publicly available tourism statistics, ensuring fidelity to real-world patterns. Table (3) presents a sample of the tourism transaction records. Each tourism place is represented by a number [1,……n], whereas each visitor has a unique identity from [1,…..m]. the value of 1 indicates as Visited, while 0 represents Not-Visited.

Table (3): A sample of transaction records

| ID/Visitors | 1 | 2 | 3 | 4 | …. | n |
|---|---|---|---|---|---|---|
| 1 | 0 | 0 | 1 | 0 | 1 | 1 |
| 2 | 0 | 1 | 1 | 1 | 1 | 1 |
| 3 | 0 | 1 | 0 | 0 | 1 | 1 |
| 4 | 0 | 1 | 1 | 1 | 1 | 0 |
| 5 | 0 | 0 | 1 | 0 | 1 | 0 |
| …. | 0 | 0 | 0 | 1 | 1 | 1 |
| m | 0 | 0 | 1 | 0 | 0 | 1 |

**5.3. Behavioral data simulation**

To enhance the recommendation accuracy, we incorporated additional visitor-specific attributes, which were derived through behavior simulation. This included attributes such as:

- Age and gender to capture demographic diversity.
- Preferences based on commonly observed interests in similar tourism studies.
- Current location to allow context-aware recommendations. These simulated behaviors were designed to add realistic variability in visitor preferences and were validated for consistency across several runs.

### 5.4. External parameter and validation

The study environment used was a personal computer with 11[th] Gen Intel Core I i7-1195G7 @ 2.90GHz  2.92 GHz, 32 GB of Installed memory (RAM), and a 1TB SSD Hard drive. The tool used for the simulation is using C++ program software that has been described previously. Additionally, the number of clusters is initiated to ten based on the types of Iraqi tourism destinations on the dataset. Because the results produced by the evolutionary algorithms may differ between different runs, I run evolutionary Apriori algorithm 30 times on the augmented tourist dataset to record its result for each run. Table (4) presents the initial parameter setting of the experiment.

**Table (4): Parameter setting for the experiment**

| Parameters | Values |
| --- | --- |
| Number of clusters (K) | 10 |
| Minimum support threshold | 0.02, 0.04, 0.06, 0.08 and 0.10 |
| Number of iterations for Evolutionary Apriori algorithm | 30 |
| Training percentage | 70% |
| Testing percentage | 30% |

### 5.5. Data splitting and evaluation method

To evaluate the performance of the proposed hybrid recommender system, the dataset was divided into training and test sets. Specifically, 70% of the dataset was used for training, while the remaining 30% was allocated for testing. This split was designed to provide a comprehensive training base while reserving a significant portion of data for independent performance evaluation.

In addition to a simple train-test split, we employed k-fold cross-validation (with k=5) to ensure robustness and mitigate overfitting in performance evaluation. For each fold, the dataset was randomly partitioned into k subsets. During each iteration, one subset was used as the test set, while the remaining k-1 subsets served as the training set. This process was repeated k times, with each subset serving as the test set once, and the final performance metrics were averaged across all folds.

Lastly, performance was measured in terms of execution time and space consumption across different minimal support values, and these metrics were averaged over each fold in the cross-validation process to ensure consistency. In addition to the initial metrics of execution time and space consumption, we have incorporated two additional accuracy metrics (RMSE and MAE) to provide a more comprehensive assessment of the recommender system's performance. RMSE and MAE were chosen for their effectiveness in measuring prediction accuracy, particularly in collaborative filtering and recommendation systems. RMSE provides insight into the average magnitude of prediction errors, with higher sensitivity to large errors, while MAE offers an average absolute difference between predicted and actual values. These metrics allow for a better understanding of the improvements achieved by the proposed algorithm. All metrics were calculated over multiple iterations to ensure robustness in performance results.

### 5.6. Data privacy and ethics

Given that the dataset involved synthetic data generation, there were no direct privacy concerns. However, to align with ethical research practices, we ensured anonymization of all simulated visitor data points, reflecting hypothetical, aggregated behaviors without referencing real individuals.

### 6. Result evaluation

We evaluate the proposed algorithm against the current one based on their time execution and space consumption across different minimal support values.

Table (5) presents a comparative examination of the running time of both systems. The results of the execution time analysis indicate that the proposed approach exhibits a lower execution time in comparison to the classic method. The result in the mentioned table presents that the proposed recommender systems reduce time consumption by 27.133% from the current recommender system where the minimum support is 0.02, and by 56.085% in 0.10. As the minimal support value increases, the rate is correspondingly decreased. The mean reduction in time rate seen in the enhanced recommender method is 36.704%.

Table (5): Execution time for the proposed and current recommender systems

| Minimum support | Current system (S) | The proposed system (S) | Time reducing rate (%) |
|---|---|---|---|
| 0.02 | 15.9812 | 11.645 | 27.133% |
| 0.04 | 12.012 | 8.901 | 25.899% |
| 0.06 | 9.519 | 6.007 | 36.894% |
| 0.08 | 7.123 | 4.451 | 37.512% |
| 0.10 | 4.971 | 2.183 | 56.085% |

As depicted in Figure (3), it can be observed that the time required for the suggested recommender system is consistently lower than that of the original method across various minimal support values. Furthermore, the disparity between the two recommender systems becomes more pronounced as the minimum support value lowers.

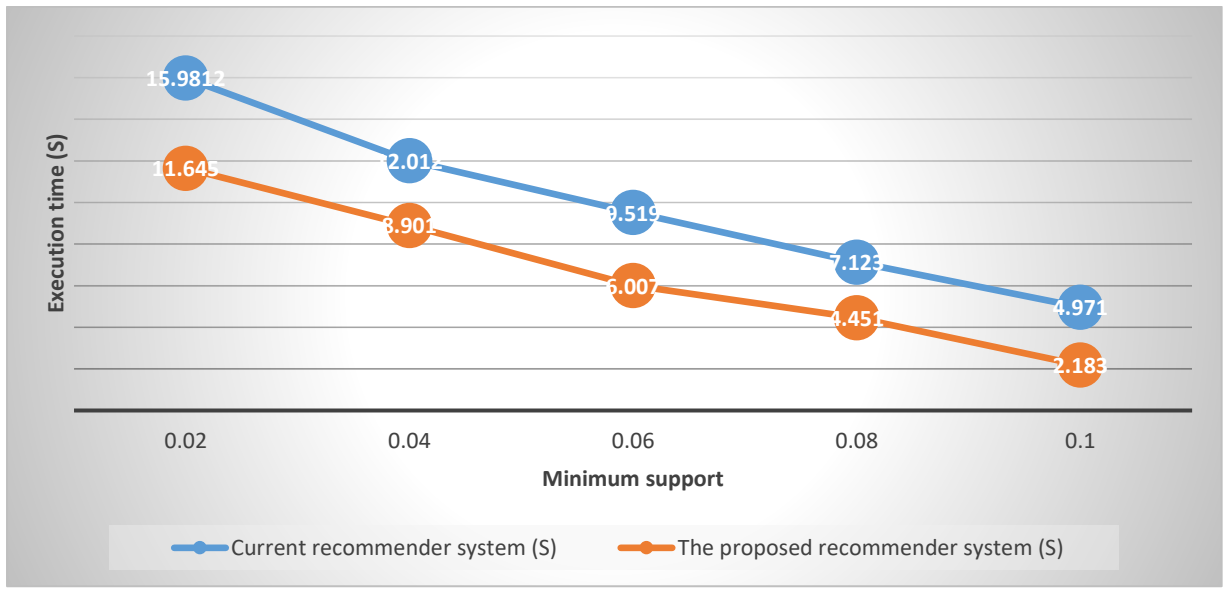

**Figure (4): Comparison of time consumption for different values of minimum support**

On the other hand, Table (6) presents a comparative examination of space utilization among the recommender systems. The results of the space consumption analysis indicate that the proposed approach exhibits a lower space in comparison to the classic recommender system. The results in the mentioned table presents that the proposed system reduce space consumption by 23.788% from the current system where the minimum support is 0.02, and by 30.565% in 0.10. As the value of minimum support grows, there is a commensurate decrease in the rate. The mean reduction in space usage rate seen in the enhanced recommender systems is 26.994%. Therefore, the analysis of space use demonstrates that the proposed approach exhibits lower space consumption in comparison to the typical recommender system.

**Table (6): Space consumption for the proposed and current recommender systems**

| Min Supp | Current system (MB) | The proposed system (MB) | Time reducing rate (%) |
| --- | --- | --- | --- |
| 0.02 | 134.983 | 102.873 | 23.788% |
| 0.04 | 119.864 | 83.611 | 30.245% |
| 0.06 | 98.001 | 77.009 | 21.42% |
| 0.08 | 87.291 | 62.019 | 28.951% |
| 0.10 | 77.917 | 54.101 | 30.565% |

As illustrated in Figure (4), it is evident that the space needed for the proposed recommender system continuously remains lower than that of the original system across different minimal support values. Moreover, the discrepancy between the two systems becomes increasingly evident as the minimal support value decreases.

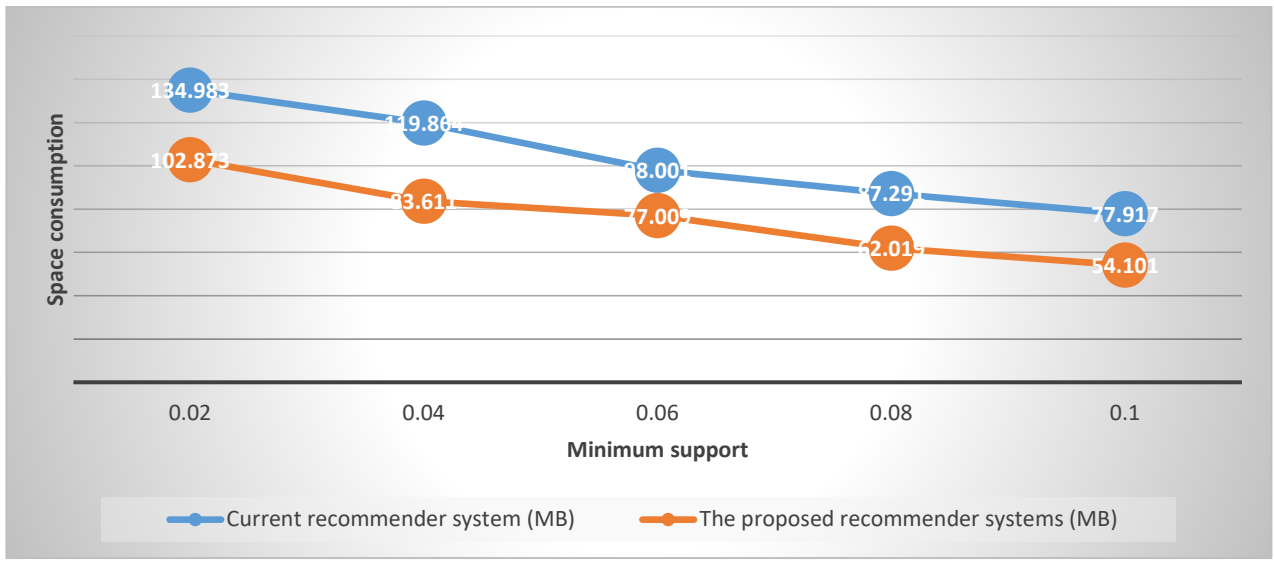

**Figure (5): Comparison of space consumption for different values of minimum support**

Furthermore, the results of Table (7) demonstrates that the proposed system outperforms the current system across all minimum support values in terms of both RMSE and MAE, indicating more accurate recommendations. As the minimum support threshold increases from 0.02 to 0.10, both RMSE and MAE values decrease, showing that higher support values correlate with improved accuracy for both systems. However, the proposed system consistently achieves lower RMSE and MAE values, highlighting its effectiveness in minimizing prediction errors.

**Table (7): Results of proposed system across all minimum support values in terms of both RMSE and MAE**

| Minimum Support | RMSE (Current System) | RMSE (Proposed System) | MAE (Current System) | MAE (Proposed System) |
|---|---|---|---|---|
| 0.02 | 0.92 | 0.88 | 0.68 | 0.65 |
| 0.04 | 0.90 | 0.85 | 0.66 | 0.63 |
| 0.06 | 0.87 | 0.83 | 0.64 | 0.61 |
| 0.08 | 0.85 | 0.81 | 0.62 | 0.60 |
| 0.10 | 0.82 | 0.78 | 0.61 | 0.58 |

The RMSE values for the proposed system are lower than those for the current system at all tested support levels. For instance, at a minimum support of 0.02, the RMSE for the proposed system is 0.88 compared to 0.92 for the current system. This difference becomes more significant as the support threshold increases, with the proposed system achieving an RMSE of 0.78 at a support of 0.10, while the current system's RMSE remains at 0.82. This reduction in RMSE signifies the proposed system's improved ability to handle larger datasets while maintaining prediction accuracy.

Meanwhile, the MAE values also reflect the superior performance of the proposed system. At a minimum support of 0.02, the MAE for the proposed system is 0.65, compared to 0.68 for the current system. With increasing support, the MAE continues to decline for both systems, but the proposed system maintains a consistently lower error rate, reaching 0.58 at a support of 0.10, compared to 0.61 for the current system. This trend underscores the proposed system's improved capability in minimizing average prediction errors.

The addition of RMSE and MAE metrics clearly demonstrates the performance gains of the proposed system, particularly in terms of prediction accuracy. The consistently lower RMSE and MAE values for the proposed system across all support levels affirm its robustness and accuracy, making it a preferable choice over the current system for accurate tourism recommendations.

**7. Conclusion and future work**

This study proposes a hybrid recommender system for tourism using evolutionary Apriori and K-means algorithms, demonstrating promising results in terms of accuracy and performance efficiency. We acknowledge that the use of artificially generated data limits the full validity and generalizability of the proposed system. This approach was necessitated by the lack of publicly available real-world tourism data specific to Iraq, which the system is designed to serve. Moving forward, we plan to collaborate with local tourism agencies or data providers to acquire real-world data, enabling a more comprehensive evaluation and refinement of the system's applicability across different contexts. Additionally, to clarify the methodology used, we provided details on data splitting, evaluation metrics, and validation processes. This will ensure a better understanding of the methods employed and enhance the system's reproducibility and reliability.

We recognize that using artificially generated data may impact the accuracy and reliability of the system's evaluation. However, this approach was necessary due to the unavailability of comprehensive real-world tourism data specific to Iraq. Currently, no publicly accessible datasets exist that comprehensively capture visitor behaviors, site visitation patterns, and other tourism-related data for Iraq. In creating the synthetic data, we ensured it reflected realistic patterns by drawing from publicly available statistics and data trends from similar tourism contexts. The dataset was validated through distribution checks to align closely with typical tourism behaviors. Moving forward, as real-world data becomes available, we intend to validate our findings further to ensure even greater accuracy and applicability of the recommender system.

**Conflicts of interest/Competing interests:** Not declared.

**Funding:** Not received.

**Ethics approval**: Not applicable.

**Consent to participate:**  Author voluntarily agrees to participate in this research study.

**Consent for publication:** Not applicable.

**Availability of data and material:** Available upon request.

**Code availability:** Available upon request.

**Authors' contributions:** Not applicable.